%% file: main.tex
\documentclass[acmlarge]{acmart}
\setcopyright{none}
\renewcommand\footnotetextcopyrightpermission[1]{}

\AtBeginDocument{%
  }

\usepackage{helvet}         
\usepackage{courier}        
\usepackage{type1cm}        
\usepackage{graphicx}        
\usepackage{multicol}        
\usepackage[bottom]{footmisc}
\usepackage{subfigure}
\usepackage{amsfonts}
\usepackage{amsmath}
\usepackage{amsfonts}
\usepackage{algpseudocode}
\usepackage{booktabs}
\usepackage{subcaption}
\usepackage{algorithm}
\usepackage{xcolor}
\usepackage{parskip}

\begin{document}

\title{Identifying and Characterising Higher Order Interactions in Mobility Networks Using Hypergraphs}

\author{Prathyush Sambaturu}
\affiliation{%
  \institution{Department of Biology \&  Pandemic Sciences Institute, University of Oxford}
  \city{Oxford}
  \country{United Kingdom}
}
\email{prathyush.sambaturu@biology.ox.ac.uk}

\author{Bernardo Gutierrez}
\affiliation{%
  \institution{Department of Biology \& Pandemic Sciences Institute, University of Oxford}
  \city{Oxford}
  \country{United Kingdom}
}
\email{bernardo.gutierrez@biology.ox.ac.uk}

\author{Moritz U.G. Kraemer}

\affiliation{%
  \institution{Department of Biology \& Pandemic Sciences Institute, University of Oxford}
  \city{Oxford}
  \country{United Kingdom}
}
\email{moritz.kraemer@biology.ox.ac.uk}

%
%


\input{abstract}

\input{ccsandkeywords}

\maketitle

\input{introduction}

\input{Preliminaries}

\input{methods}

\input{results}

\input{discussion}

\input{conlusions}

\section*{Data, Materials, and Software Availability}
All code used in this work is openly available from GitHub at \url{https://github.com/prathyushsambaturu/Covisitation-hypergraphs.git}. Data used in this work are publicly available from Yabe et al. 2024 \cite{Yabe_YJMob100K_2024}.

\section*{Acknowledgments}
M.U.G.K. acknowledges funding from The Rockefeller Foundation (PC-2022-POP-005), Google.org, the Oxford Martin School Programmes in Pandemic Genomics (\& B.G.) \& Digital Pandemic Preparedness, European Union's Horizon Europe programme projects MOOD (\#874850) and E4Warning (\#101086640), Wellcome Trust grants 303666/Z/23/Z, 226052/Z/22/Z (\& B.G.) \& 228186/Z/23/Z, the United Kingdom Research and Innovation (\#APP8583), the Medical Research Foundation (MRF-RG-ICCH-2022-100069), UK International Development (301542-403), the Bill \& Melinda Gates Foundation (INV-063472) and Novo Nordisk Foundation (NNF24OC0094346). The contents of this publication are the sole responsibility of the authors and do not necessarily reflect the views of the European Commission or the other funders.

\section*{Author contributions}
P.S., B.G., and M.U.G.K. conceptualised the study. P.S. performed the research and analysed the data. M.U.G.K. supervised the work. P.S. wrote the initial manuscript with critical input from M.U.G.K. and B.G. All authors edited and revised the manuscript. M.U.G.K. acquired funding for the research.

\bibliographystyle{unsrt}
\bibliography{refs}
\clearpage
\input{appendix}

\end{document}

%% file: abstract.tex
\begin{abstract}
Understanding human mobility is essential for applications ranging from urban planning to public health. Traditional mobility models such as flow networks and colocation matrices capture only pairwise interactions between discrete locations, overlooking higher-order relationships among locations (i.e., mobility flow among two or more locations). To address this, we propose co-visitation hypergraphs, a model that leverages temporal observation windows to extract group interactions between locations from individual mobility trajectory data. Using frequent pattern mining, our approach constructs hypergraphs that capture dynamic mobility behaviors across different spatial and temporal scales. We validate our method on a publicly available mobility dataset and demonstrate its effectiveness in analyzing city-scale mobility patterns, detecting shifts during external disruptions such as extreme weather events, and examining how a location's connectivity (degree) relates to the number of points of interest (POIs) within it. Our results demonstrate that our hypergraph-based mobility analysis framework is a valuable tool with potential applications in diverse fields such as public health, disaster resilience, and urban planning.
\end{abstract}

%% file: ccsandkeywords.tex
\ccsdesc[300]{Mathematics of computing~Hypergraphs}
\ccsdesc[300]{Information systems~Data mining}
\ccsdesc[300]{Applied computing}

\keywords{hypergraphs, higher-order interactions, frequent pattern mining, human mobility data}

%% file: introduction.tex
\section{Introduction}
Human mobility data is crucial for understanding patterns of movement across geographical regions, with applications spanning urban planning\cite{ghahramani_urban_2020}, transportation systems design\cite{renzo_appsciences_2020}, infectious disease modeling and control \cite{Kraemer2020-sk, Belik2011-wr}, and social dynamics studies \cite{Zhao2016}. Traditionally, mobility data has been represented using flow networks\cite{wang_cikm_2017, yeghikyan_smartcomp_2020} or colocation matrices \cite{IYER2023100663}, where the primary representation is via pairwise interactions. In flow networks, this means directed edges represent the movement of individuals between two locations; colocation matrices measure the probability that a random individual from a region is colocated with a random individual from another region at the same location. These data types and their pairwise representation structure have been used to identify the spatial scales and regularity of human mobility, but have inherent limitations in their capacity to capture more complex patterns of human movement involving higher-order interactions between locations -- that is,  group of locations that are frequently visited by many individuals within a period of time (e.g., a week) and revisited regularly over time. Higher-order interactions between locations can contain crucial information under certain scenarios. For example, in contact tracing during the early stages of an epidemic, identifying these higher-order interactions of locations over a period of, say, three days enables health authorities to trace all locations visited by potentially exposed individuals during that time. Pairwise representations of mobility data are not useful for this type of analysis. Addressing these limitations requires a more expressive representation of mobility data.

Unlike simple graphs, which model pairwise interactions, hypergraphs~\cite{springerHypergraphTheory} can represent higher-order relationships, where a single edge (hyperedge) connects an arbitrary number of nodes. Hypergraphs have been employed in various domains, such as epidemic modeling\cite{Higham2021-qy, Bodo2016-ua} and biological systems\cite{schowb_bio_2021}, where higher-order interactions between entities play a significant role in the spread, maintenance, and stability of these systems. 

In this paper we propose a novel approach to construct \textit{co-visitation hypergraphs} from individual-level mobility trajectory data such as those available from eXtended Data Records (XDR) or GPS traces\cite{pappalardo2021}. Our method captures higher-order interactions of locations while allowing the study of mobility dynamics across varying temporal and spatial scales through a parameter called the \textit{observation window length} ($\Delta T$). Temporal windows divide the timeline into overlapping segments, allowing insights into how group interactions evolve over daily or weekly periods. We also develop analyses to understand the spatio-temporal mobility behaviors from the constructed hypergraphs. Our framework facilitates a detailed examination of the structural and spatial properties of hypergraphs and their change across temporal windows and during periods characterised by business-as-usual behaviour as well as external shocks. 

Recent studies relevant to hypergraph-based mobility analysis focus mainly on hypergraph embedding approaches ~\cite{Yang2019-ps}\cite{Zhang-hypergraph_rep_2024}. Yang et al. 2019~\cite{Yang2019-ps} embed user mobility and social relationships using hypergraphs, linking users, locations, times, and activities to perform tasks such as friendship and location prediction. In contrast, our work focuses on locations rather than users, constructing co-visitation hypergraphs to capture higher-order interactions between locations visited together within specific temporal windows ($\Delta T$). This location-centric perspective is particularly useful for analyzing changing mobility behaviors across regions and during emergencies, such as natural disasters or infectious disease outbreaks.

We summarize our contributions in this work as follows:
\begin{itemize}
    \item \textbf{Temporal window-based co-visitation hypergraph}: We introduce a novel hypergraph-based representation of mobility data that captures complex patterns of human movement from a location-centric perspective.  
    \item \textbf{Algorithm to construct co-visitation hypergraph}: 
   We introduce a flexible algorithm to construct co-visitation hypergraphs from individual-level mobility data using temporal observation windows ($\Delta T$), which can enable analysis of co-visitation patterns over varying time scales.
   \item \textbf{Application to individual-Level mobility data}:
    We validate our approach using YJMob100K\cite{Yabe_YJMob100K_2024} dataset performing the following analysis.
    \begin{itemize}
        \item  \textit{Structural and spatial analysis}. We analyze the structural and spatial properties of co-visitation hypergraphs to characterize each location's role in group interactions, determining whether it is central or peripheral and whether it is co-visited with nearby or distant locations. Further, we investigate how our algorithmic parameters influence key hypergraph characteristics. Specifically, we investigate hyperedge size distributions, spatial coverage, and the evolution of higher-order interactions. 
        \item \textit{Framework to examine mobility changes during external shocks}. We extend the analysis to study changes in mobility patterns during emergency scenarios using a pre-defined phase in the YJMob100K dataset. By comparing ``regular'' and ``emergency'' days, we identify significant differences in hyperedge structures, spatial coverage, and interaction patterns. This framework demonstrates the utility of co-visitation hypergraphs for analyzing disruptions in mobility, such as those caused by extreme weather events or public health crises, offering insights into shifts in co-visitation dynamics under such conditions.
    \end{itemize}
\end{itemize}

Our framework has potential applications in studying mobility behaviors during emergencies, disasters, and provide the potential interaction network for modelling infectious diseases. Further, outputs could be used to understand the evolution of co-visitaiton hypergraphs, how different locations emerge as central nodes and those that lose their centrality.

%% file: preliminaries.tex
\section{Preliminaries}
In this section, we provide the necessary background to clarify the rest of the paper.

\subsection{A Brief Overview of Hypergraphs}
\begin{figure}
\centering
\includegraphics[scale=0.45]{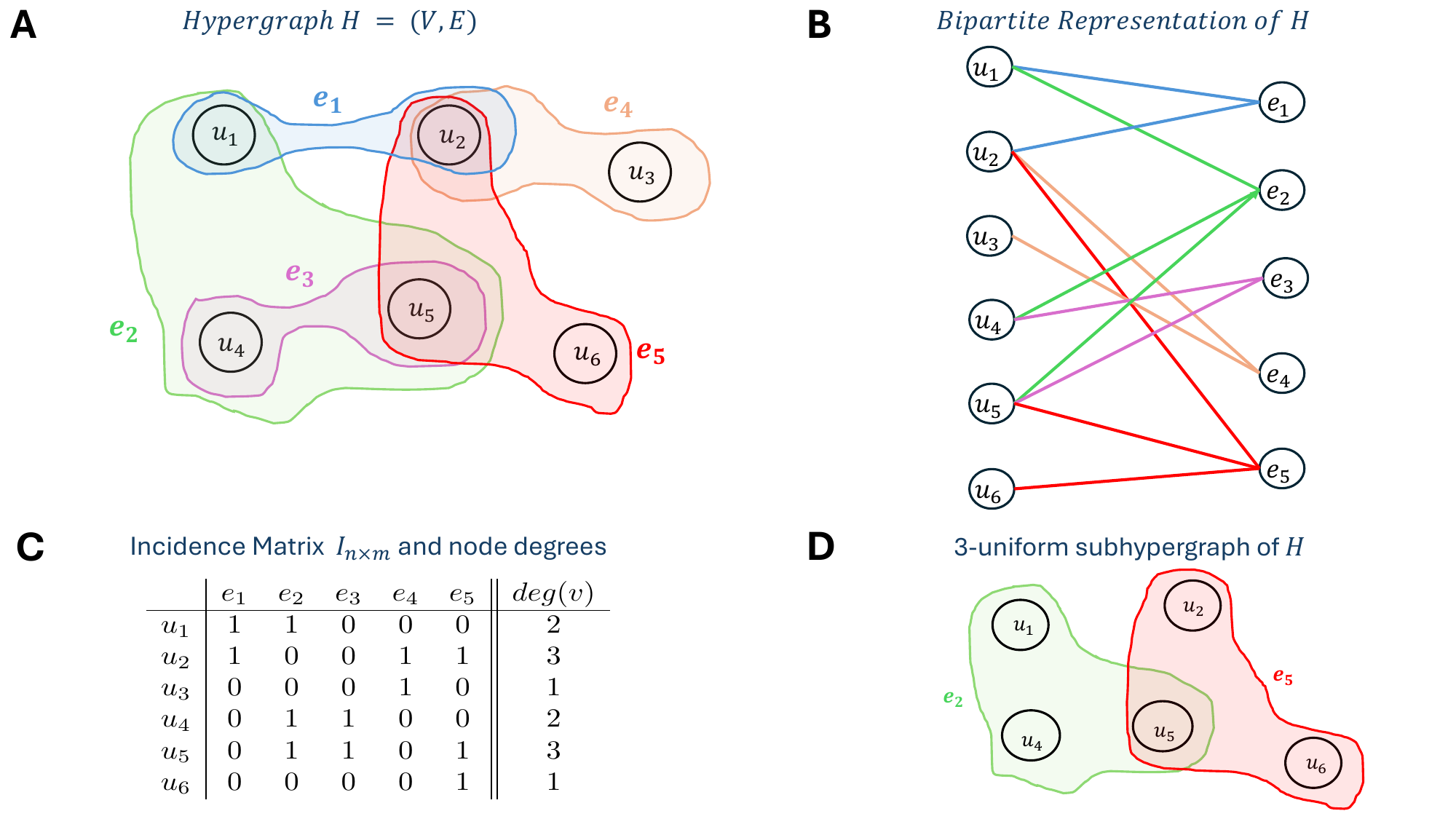}
    \caption{\textit{Hypergraph Example.}  
    \textbf{A}) The hypergraph $H = (V, E)$ consists of 6 vertices and 5 hyperedges. The vertex set is $V = \{u_1, u_2, u_3, u_4, u_5, u_6\}$ and the hyperedge set is $E = \{\{e_1 = \{u_1, u_2\}, e_2 = \{u_1, u_4, u_5\}, e_3 = \{u_4, u_5\}, e_4 = \{u_2, u_3\}, e_5 = \{u_2, u_5, u_6\}\}$. The vertices are represented as circles, and the hyperedges are depicted as colored regions encompassing the vertices involved in each hyperedge.
    \textbf{B}) The corresponding bipartite graph of $H$ is shown, where $V$ forms one partition and $E$ forms the other. The edges in the bipartite graph are colored according to the hyperedge colors from the original hypergraph.  
    \textbf{C}) The incidence matrix $I = \{I_{ij}\}$ is shown, where $I_{ij} = 1$ if hyperedge $e_j$ is incident on vertex $u_i$. The matrix is accompanied by the vertex degrees, calculated as the sum of the entries in each row. The vertices $u_2$ and $u_5$ have the highest degree (3), while $u_3$ and $u_6$ have the lowest degree (1).  
    \textbf{D}) The 3-uniform subhypergraph of $H$ includes two hyperedges of cardinality 3: $e_2 = \{u_1, u_4, u_5\}$ and $e_5 = \{u_2, u_5, u_6\}$, with their intersection being the vertex $u_5$.
}
\label{fig:hypergraph_example}
\end{figure}

A hypergraph is denoted by $H = (V, E)$, where $V = \{u_1, \cdots, u_n\}$ is a set of vertices (or nodes), and $E$ is a family of subsets of $V$ called hyperedges. A hyperedge $e_j = \{u_{j1}, \dots, u_{jl}\} \in E$ generalizes the concept of an edge in a simple graph, where $|e_j| = l$ for $l \geq 0$. In the special case of a simple graph, each edge has exactly two vertices ($l = 2$). Hypergraphs, however, allow for higher-order interactions among entities, unlike simple graphs which are limited to pairwise interactions. Figure \ref{fig:hypergraph_example}A shows an example hypergraph with six vertices (or nodes) and five hyperedges.

A hypergraph $H = (V, E)$ can be represented as a bipartite graph $B = (V \cup E, \; E_B)$, with two partitions: the set of vertices $V$ in one partition and the set of hyperedges $E$, which are represented as nodes in the second partition. In the bipartite graph $B$, an edge $(u_i, e_j) \in E_B$ exists if and only if the hyperedge $e_j$ in $H$ is incident on the vertex $u_i$. The bipartite representation of the hypergraph from Figure \ref{fig:hypergraph_example}A is illustrated in Figure \ref{fig:hypergraph_example}B.

The \textit{incidence matrix} of a hypergraph $H$ is denoted by $I_{n \times m} = \{I_{ij}\}$ ($m$ denotes number of hyperedges in $H$), where $I_{ij} = 1$ if hyperedge $e_j$ is incident on node $u_i$ in $H$, and $I_{ij} = 0$ otherwise. The \textit{degree} of a vertex, $u_i \in V$, is the number of hyperedges incident on that vertex, calculated as $\text{deg}(u_i) = \sum_{j} I_{ij}$. Figure\ref{fig:hypergraph_example}C shows the incidence matrix and the degree of each node vertex for the example hypergraph.

The size of a hyperedge is defined as the number of vertices it contains. For example, the size of hyperedge $e_2 = \{u_1, u_4, u_5\}$ in the hypergraph $H$ shown in Figure \ref{fig:hypergraph_example} is 3. The rank $r(H)$ of a hypergraph is the maximum size of any hyperedge in $H$. In the example shown in Figure \ref{fig:hypergraph_example}, the hypergraph $H$ has three hyperedges ($e_1, e_3$, and $e_4$) of size 2, and two hyperedges ($e_2$ and $e_5$) of size 3. Therefore, the rank of the hypergraph $H$, $r(H)$, is 3, since the largest hyperedges consists of three vertices. 

A $k$-uniform hypergraph is a hypergraph in which all hyperedges have size $k$. A $k$-uniform subhypergraph of a hypergraph $H$ is a subhypergraph induced by all hyperedges in $H$ of size exactly $k$. Specifically, for a 3-uniform subhypergraph $H' = (V', E')$, we have $E' \subseteq E$, where $E' = \{ e \in E : |e| = 3 \}$. The 3-uniform subhypergraph of our example hypergraph (shown in Figure\ref{fig:hypergraph_example}D) contains only two hyperedges $e_2 = \{u_1, u_4, u_5\}$ and $e_5 = \{u_2, u_5, u_6\}$ which intersect at vertex $u_5$.
\subsection{Existing Representations of Human Mobility Based on Location-to-Location Interactions}
Empirical human mobility patterns can be inferred from various sources, such as mobile phones, which is commonly available in the form of Call Detail Records (CDR), eXtended Detail Records (XDR), or GPS traces \cite{pappalardo2021}. XDR and GPS trace data typically offer detailed mobility trajectories for individuals within a specific geographical region along with corresponding timestamps. This means that not only pairwise exchanges between locations can be inferred by the sequential visitation of locations, including the approximate time of stay in each. 

A mobility flow network is a weighted directed graph where nodes represent locations and directed edges indicate the movement of individuals from one node to another, and edge weights quantify the magnitude of that movement. However, this representation cannot accurately capture complex flows, namely, movements of individuals among more than two locations (see Figure \ref{fig:flownet_implication} in Appendix).

Colocation maps~\cite{IYER2023100663} estimate the likelihood that individuals from different home locations are present in the same place at the same time, as illustrated by a colocation event in Figure \ref{fig:flownet_implication}B in Appendix. But they do not have information about where the colocations occur. These maps reveal how populations from various locations come into contact, emphasizing the spatial and temporal overlap of individuals within a given time window. However, this approach is limited to pairwise relationships between locations and does not account for the dynamic interactions or movement flows between them.

Chang et al. 2021 \cite{Chang2021-md} proposed a bipartite network representation where neighbourhoods (census block groups) form one set of nodes and point-of-interests (POIs, such as restaurants, shopping centers) form another set of nodes. They incorporate this representation in epidemic modelling to predict inequities faced among different groups of people during COVID-19.

All the above existing location-to-location mobility data representations solely focus on pairwise interactions.

%% file: methods.tex
\section{Methods}
In this section, we present steps in our methodology for constructing a co-visitation hypergraph from individual-level mobility data. Figure \ref{fig:algo_schematic} provides a schematic overview of the key steps in the process formalized in the pseudocode of the Algorithm \ref{alg:mobility_hypergraph}. 

\subsection{Data Preprocessing and Notations}
Let the region of interest be partitioned into a set of locations $\mathcal{L} = \{\ell_1, \ell_2, \dots, \ell_n\}$ that collectively cover the entire geographical region of interest. Define $[D] = \{1, 2, \dots, D\}$ as the set of days and $\mathcal{U} = \{u_1, \dots, u_r\}$ as the set of individuals. The mobility trajectories are represented by a function $\mathbb{D} : \mathcal{U} \times [D] \rightarrow 2^{\mathcal{L}}$, mapping a pair of an individual and a day to a subset of locations in $\mathcal{L}$. Specifically, $\mathbb{D}(u_i, d)$ denotes the set of locations visited by individual $u_i$ on day $d$, disregarding the sequence and frequency of visits to each location. 

We pre-process the individual-level mobility data to construct $\mathbb{D}$. Figure \ref{fig:algo_schematic}A provides an example of sample trajectories for three individuals, $u_1, u_2$, and $u_3$, shown in both raw format and the resulting set-based notation. While the unit of time is assumed to be one day in this section, this can be adjusted by redefining $\mathbb{D}$ accordingly.

\subsection{Observation Windows}
\begin{figure*}    \centering
    \includegraphics[height=9cm, width=16cm]{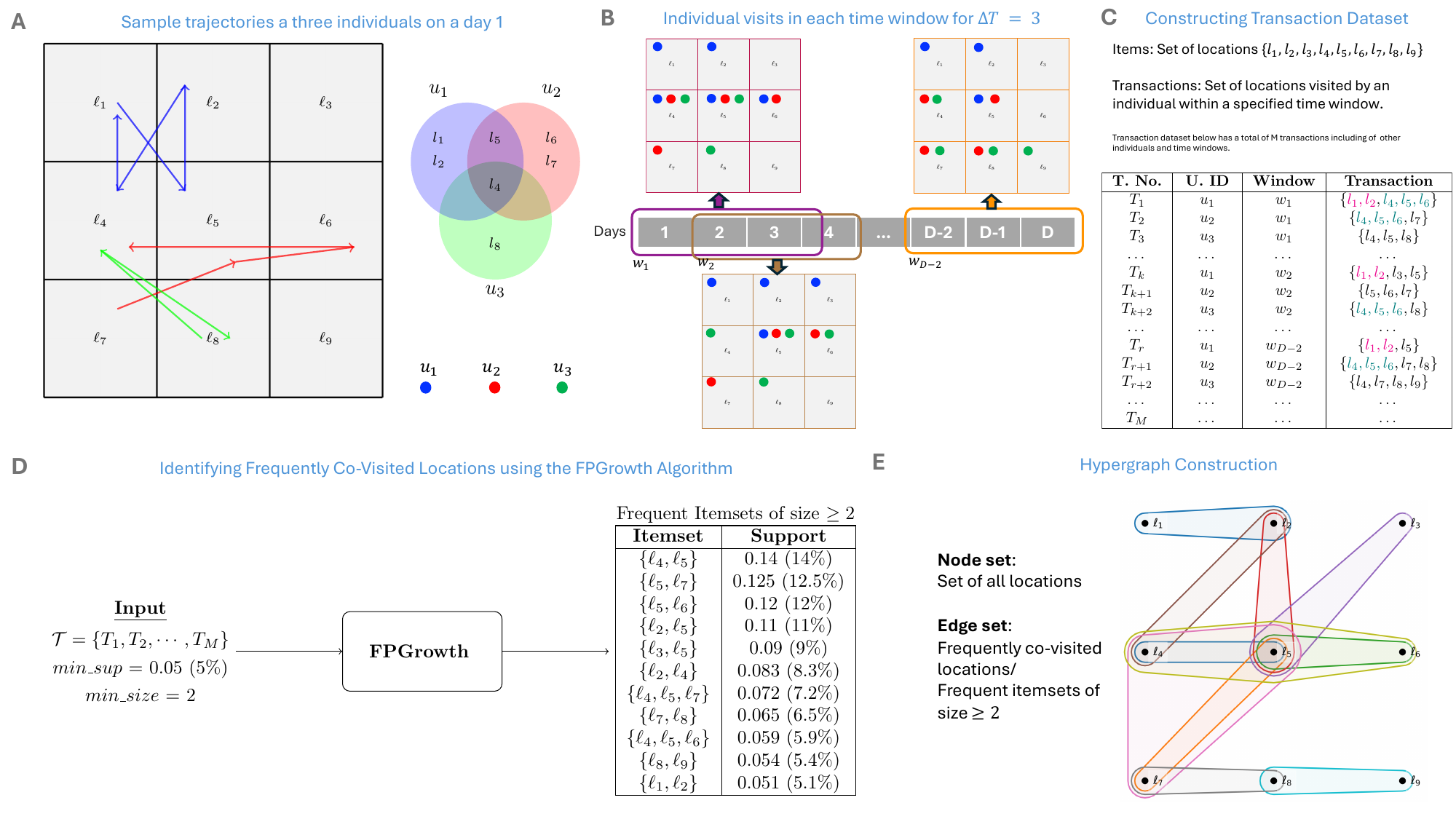}
    \caption{
    \textit{Schematic of constructing a co-visitation hypergraph from individual-level human movement data.} \textbf{A}) (left) Sample trajectories for three individuals, $u_1$, $u_2$, and $u_3$, over a single day, shown on a 3x3 grid map in blue, red, and green. (right) These trajectories are summarized as sets of visited locations: $\{\ell_1, \ell_2, \ell_4, \ell_5\}$ for $u_1$, $\{\ell_4, \ell_5, \ell_6, \ell_7\}$ for $u_2$, and $\{\ell_4, \ell_8\}$ for $u_3$.
     \textbf{B}) A sliding time window of length $\Delta T = 3$ days moves across a sequence of $D$ days, creating observation windows ${w_1, w_2, \dots, w_{D-2}}$. For representative windows $w_1$, $w_2$ shown, and $w_{D-2}$, each circle on the grid represents a visit by one of the three individuals to a location, with colors indicating which individual visited (blue for $u_1$, red for $u_2$, and green for $u_3$). \textbf{C}) In each observation window, the locations visited by an individual are treated as a transaction, with the locations forming the set of items. This converts mobility data into a transaction dataset $\mathcal{T} = \{T_1, \cdots, T_M\}$, allowing the identification of frequently co-visited locations across $M$ individuals over $D$ days. Potential candidates for frequently co-visited locations (frequent itemsets) are highlighted: \textcolor{magenta}{$\{\ell_1, \ell_2\}$} and \textcolor{teal}{$\{\ell_4, \ell_5, \ell_6\}$}.
     \textbf{D}) The FPGrowth algorithm takes the transaction dataset $\mathcal{T}$, a minimum support threshold $min\_sup$ (set to 0.05 in this example, meaning an itemset must appear in at least 5\% of transactions to be considered frequent), and a minimum itemset size $min\_size$ (set to 2) as inputs. It identifies all frequent itemsets that meet these criteria. The table on the right displays the results for the example, showing 9 frequent itemsets of size 2 and 2 frequent itemsets of size 3. \textbf{E}) The mobility hypergraph is constructed from the frequent itemsets identified in the previous step, where locations serve as nodes and frequent itemsets form the hyperedges. In this example, location $\ell_5$ has the highest degree (7), while locations $\ell_1$, $\ell_3$, and $\ell_9$ have the lowest degree (1). Both hyperedges of size 3 are incident on $\ell_5$.
     }
    \label{fig:algo_schematic}
\end{figure*}

To identify sets of locations frequently co-visited by individuals, we segment time into overlapping observation windows of length $\Delta T$. These windows are generated by applying a sliding window of length $\Delta T$ over the $D$ days, resulting in a sequence of observation windows $\{w_1, \dots, w_{D-\Delta T +1}\}$. The $t^{th}$ observation window, $w_t$, encompasses the days $\{t, t+1, \dots, t+\Delta T -1\}$ (see Figure \ref{fig:algo_schematic}B).

For each observation window, we aggregate the set of locations visited by an individual during the window, ignoring the frequency and order of visits. This approach ensures we focus on co-visited locations without considering temporal details within the window. Figure~\ref{fig:algo_schematic}B illustrates the concept of observation windows.

\subsection{Identifying Frequently Co-visited Locations Using Frequent Pattern Mining}
We use frequent pattern mining techniques~\cite{Agrawal1994-kk, Han2000-nb} to identify sets of locations frequently co-visited by individuals in the mobility data. Frequent pattern mining involves discovering frequent itemsets in a transactional database, where each transaction consists of a subset of all items.

To apply this methodology, we first prepare a transactional dataset. For an individual $i$, the set of locations visited during an observation window $w_t = \{t, \cdots, t+\Delta T -1\}$ is defined as:
\begin{equation*}
T_j = \bigcup_{d=t}^{t+\Delta T-1} \mathbb{D}(i, d),
\end{equation*}
where $\mathbb{D}(i, d)$ represents the set of locations visited by individual $i$ on day $d$.

We treat each $T_j$ as a single transaction. Thus, for each individual and each observation window, we create one transaction, resulting in up to $M \leq (D - \Delta T + 1)r$ transactions, where $D$ is the total number of days, $\Delta T$ is the window size, and $r$ is the number of individuals. The resulting transaction dataset is denoted as $\mathcal{T} = \{T_1, \cdots, T_M\}$. Figure \ref{fig:algo_schematic}C shows an example of such a transaction dataset.

An \textit{itemset}, denoted by $e \subseteq \mathcal{L}$, is a subset of locations (items) that can appear in any transaction in the dataset. The support of an itemset $e$, denoted by $\mathbf{S}(e)$, is the proportion of transactions in $\mathcal{T}$ that contain $e$. An itemset is considered \textit{frequent} if its support exceeds a user-defined threshold, $min\_sup$, which specifies the minimum required proportion of transactions containing the itemset. For example, setting $min\_sup = 0.05$ means an itemset is frequent if it appears in at least $5\%$ of the total transactions, or equivalently, in at least $0.05M$ transactions.

A frequent itemset $e \subseteq \mathcal{L}$ is called a \textit{maximal frequent itemset} if no superset $e' \supset e$ is frequent. In other words, $e$ is frequent and adding any additional items to $e$ results in an itemset that does not meet the minimum support threshold.

We apply the FP-Growth algorithm~\cite{Han2000-nb} to compute frequent itemsets from our transaction dataset $\mathcal{T}$. The algorithm requires two inputs: the minimum support threshold, $min\_sup$, and the minimum size of an itemset, $min\_size$. FP-Growth first constructs a compact \textit{FP-tree} (Frequent Pattern Tree) from the transaction data, encoding the frequency of itemsets while preserving their hierarchical relationships. It then recursively mines the FP-tree to extract  frequent location patterns $\mathcal{F}$ and their corresponding supports $\mathbf{S}$.

In this context, the frequent itemsets represent sets of locations co-visited by individuals with high support within observation windows of length $\Delta T$. These patterns provide insights into mobility behaviors in the population and highlight significant co-visitation trends.

\subsection{Constructing the Co-visitation Hypergraph}
We define the co-visitation hypergraph as $\mathbb{H} = (V, \mathcal{E}, w)$, where $V = \mathcal{L}$ represents the set of locations, $\mathcal{E}$ denotes the set of hyperedges, and $w: \mathcal{E} \rightarrow \mathbb{R}$ is a weight function that assigns a weight $w(e)$ to each hyperedge $e \in \mathcal{E}$. For each frequent itemset $e \in \mathcal{F}$, a corresponding hyperedge is added to $\mathcal{E}$. The weight of a hyperedge $e$ is defined as the support of the corresponding itemset $e$, given by $w(e) = \mathbf{S}(e)$.

The resulting hypergraph $\mathbb{H}$ (example in Figure \ref{fig:algo_schematic}E) encodes the relationships among frequently co-visited locations, with the weights of the hyperedges reflecting the frequency of co-visitation.

\begin{algorithm}[H]
\caption{Constructing a Co-Visitation Hypergraph from Mobility Data}
\label{alg:mobility_hypergraph}
\begin{algorithmic}[1]
\Require Locations $\mathcal{L}$, Mobility data $\mathbb{D}$ for $r$ individuals over $D$ days, Time window $\Delta T$, $min\_sup$, $min\_size$
\Ensure Hypergraph $\mathbb{H} = (\mathcal{L}, \mathcal{E}, w)$

\State Initialize $\mathcal{T} \gets \emptyset$, $\mathbb{H} \gets (\mathcal{L}, \emptyset)$ \Comment{Initialize dataset and hypergraph}

\For {each window $w_t$ of length $\Delta T$ over $D$ days} \Comment{Slide observation window}
    \For {each individual $i \in \{1, \dots, r\}$} \Comment{Process movements for each individual}
        \State Compute $T_j = \bigcup_{d=t}^{t+\Delta T-1} \mathbb{D}(i, d)$ \Comment{Group locations visited by individual in window}
        \If {$T_j \neq \emptyset$} add $T_j$ to $\mathcal{T}$ \Comment{Add non-empty transaction}
        \EndIf
    \EndFor
\EndFor

\State $\mathcal{F}, \mathbf{S} \gets \text{FPGrowth}(\mathcal{T}, min\_sup, min\_size)$ \Comment{Mine frequent itemsets and supports}

\For {each $e \in \mathcal{F}$} \Comment{Construct hypergraph edges with support as edge weights}
    \State Add hyperedge $e$ to $\mathcal{E}$ and set $w(e) \gets \mathbf{S}(e)$
\EndFor

\State \Return $\mathbb{H}$ \Comment{Return the constructed hypergraph}
\end{algorithmic}
\end{algorithm}

%% file: results.tex
\section{Results}

\subsection{Dataset Overview and Experiment Setup}
This section provides an overview of the datasets used in our study and outlines the experimental setup.

\subsubsection{Datasets}
We use two public, city-scale, individual-level human mobility datasets \cite{Yabe_YJMob100K_2024} to construct mobility hypergraphs, analyze their structural properties, and investigate mobility patterns. Both datasets contain movement records over 75 days with fields: user ID (\textit{uid}), day (0 to 74), timeslot (30-minute intervals labeled 0 to 47), and x/y coordinates mapped to a 200x200 grid, where each grid cell covers a 500-meter square area.

The first dataset, \textit{DS1}, captures the movements of 100,000 individuals over 75 days. The second dataset, \textit{DS2}, includes movements of 25,000 individuals during 60 regular days, followed by 15 days during an emergency scenario. No information about the nature of the emergency scenario is provided. In addition, the data sets provide the number of points of interest (POIs) by category for each location. Detailed dataset preparation can be found in Yabe et al. 2024 \cite{Yabe_Nature_2024}.

\subsubsection{Selection of spatial resolution and parameters}  
The application of our method to the above datasets involves making choices regarding spatial resolution and parameter selection ($\Delta T$, \textit{min\_sup}) as outlined below:  
\begin{enumerate}
    \item \textbf{Spatial Resolution}: The original 200x200 grid is aggregated into a 20x20 grid using a scaling factor of 10, where each cell represents a 25-km$^2$ area.  
    \item \textbf{Temporal Window ($\Delta T$)}: We examine three temporal resolutions: $\Delta T \in \{1, 3, 7\}$ days, corresponding to daily, three-day, and weekly observation windows.   
    \item \textbf{Support Threshold (\textit{min\_sup})}: We explore support thresholds \textit{min\_sup} $\in \{0.005, 0.01, 0.015\}$, corresponding to 0.5\%, 1\%, and 1.5\% minimum supports in the transactional dataset.  
\end{enumerate}  

\begin{table*}[h]
    \centering
    \caption{Experimental Parameters and Their Descriptions}
    \begin{tabular}{|l|c|l|}
        \hline
        \textbf{Parameter} & \textbf{Values} & \textbf{Description} \\
        \hline
        Spatial scaling factor & 10 & Aggregates the grid to 20x20, where each cell covers 25 km$^2$ \\
        Temporal window size ($\Delta T$) & $\{1, 3, 7\}$ days & Daily, three-day, and weekly observation windows \\
        Support threshold (\textit{min\_sup}) & \{0.005, 0.01, 0.015\} & Minimum support thresholds: 0.5\%, 1\%, and 1.5\% \\
        \hline
    \end{tabular}
    \label{tab:experiment_params}
\end{table*}  

\noindent
Table \ref{tab:experiment_params} summarizes the experimental parameters. This design facilitates a comprehensive analysis of co-visitation patterns and mobility hypergraph structures under varying temporal resolutions and support thresholds.  

\subsection{Impact of $min\_sup$ and $\Delta T$ on the Spatial Spread of Co-Visitation Hypergraphs}

We explore how the time window size ($\Delta T$) and the minimum support threshold ($min\_sup$) influence the structural and spatial characteristics of co-visitation hypergraphs. Figures \ref{fig:impact_deltaT_minsup}A-B (in Appendix) illustrate co-visitation hypergraphs for daily ($\Delta T = 1$) and three-day ($\Delta T = 3$) observation windows, respectively. For a fixed $min\_sup$ threshold, increasing $\Delta T$ results in a significant increase in the number and spatial coverage of hyperedges reflecting how a longer temporal window captures a more diverse set of interactions across the spatial grid.

Figures \ref{fig:impact_deltaT_minsup}C-D (in Appendix) provide visualizations of 5-uniform subhypergraphs derived from the co-visitation hypergraphs for $\Delta T = 3$ and $\Delta T = 7$, respectively, with $min\_sup = 0.005$.

We compute maximum chebyshev distance between nodes in each hyperedge to examine the spatial proximity of higher-order interactions. The chebyshev distance\cite{De_Oliveira2020-ay} between two nodes at coordinates $(x_1, y_1)$ and $(x_2, y_2)$ is defined as $d_\infty((x_1, y_1), (x_2, y_2)) = \max(|x_1 - x_2|, |y_1 - y_2|)$. For each hyperedge $e$, the maximum chebyshev distance is given by $d_\infty(e) = \max_{u, v \in e} d_\infty(u, v)$, where $u$ and $v$ are any two nodes in the hyperedge $e$. Therefore, the maximum chebyshev distance in a hypergraph $H$ as $D_\infty(H) = \max_{e \in \mathcal{E}} d_\infty(e)$. For $\Delta T = 3$, the maximum chebyshev distance in the 5-uniform subhypergraph is 3, while for $\Delta T = 7$, it increases to 4. This demonstrates how larger temporal windows not only lead to more higher-order interactions but also allow interactions to occur over greater spatial distances.

Visualising the hypergraphs, we observe that increasing $\Delta T$ leads to a more diffuse structure, capturing interactions that span broader spatial regions. Simultaneously, varying $min\_sup$ alters the density and strength of interactions within the hypergraph, with lower thresholds enabling a richer representation of co-visitation patterns. These findings suggests how the choice of parameter will reflect in the emergence of distinct patterns in the co-visitation hypergraph.
\subsection{Structural and Spatial Properties of the Co-visitation Hypergraph}

This section analyzes the structural and spatial characteristics of the co-visitation hypergraph by examining degree distributions, hyperedge sizes, spatial arrangements, and higher-order interactions. 

To characterize the degree distribution, we examine the complementary cumulative distribution function (CCDF)\cite{Sala_ccdf_2011} of node degrees in a hypergraph $\mathbb{H} = (V, \mathcal{E})$, denoted as $P(\text{Degree} \geq k)$, which represents the probability that a randomly selected node has a degree of at least $k$ defined as below:
\begin{equation*}
    P(\text{Degree} \geq k) = \frac{|\{v \in V : \text{Degree}(v) \geq k\}|}{|V|},
\end{equation*} 
where $|V|$ is the total number of nodes in the hypergraph.

\begin{figure}
    \centering
    \includegraphics[scale=0.47]{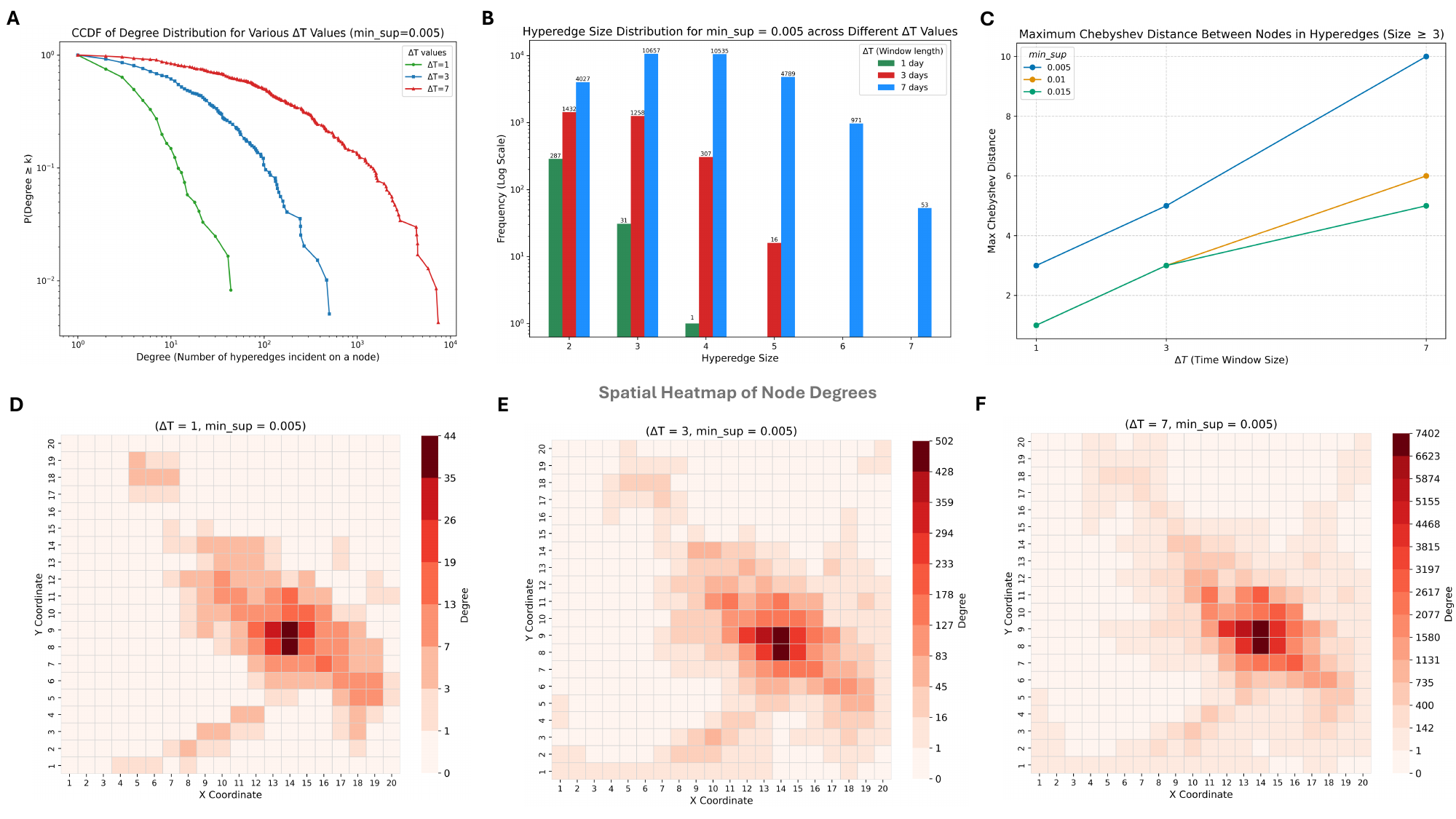}
    \caption{\textit{Structural and Spatial Properties of the Co-Visitation Hypergraph.}  
\textbf{A)} CCDF of node degrees for hypergraphs with $min\_sup = 0.005$ and $\Delta T \in \{1, 3, 7\}$ (green, blue, and red, respectively). The x-axis shows node degree, and the y-axis shows $P(\text{Degree} \geq k)$.  
\textbf{B)} Hyperedge size distribution for $\Delta T \in \{1, 3, 7\}$, with the x-axis representing hyperedge size and the y-axis representing frequency in log-scale. The maximum hyperedge size is 4 for $\Delta T = 1$, and 5 and 7 for $\Delta T = 3$ and $\Delta T = 7$.  
\textbf{C)} Maximum chebyshev distance over $\Delta T$ values. X-axis represents $\Delta T$ and y-axis represents the max chebyshev distance. Each curve corresponds to a particular $min\_sup$ value.
\textbf{D,E,F)} Spatial heatmaps of node degrees for hypergraphs with $min\_sup = 0.005$ and $\Delta T \in \{1, 3, 7\}$ on a 20x20 grid. Nodes are represented as grid cells according to their $(x, y)$ coordinates, with darker shades of red indicating higher node degrees. 
}
    \label{fig:structural_properties}
\end{figure}

Figure \ref{fig:structural_properties}A shows the CCDF of node degrees for hypergraphs with $min\_sup = 0.005$ and $\Delta T \in \{1, 3, 7\}$. While the curves exhibit a heavy tail, the degree distribution is better modeled by an exponential distribution rather than a power-law. Statistical fitting yielded log-likelihood ratios favoring an exponential distribution over a power-law for all $\Delta T$ values, with the best fit given by:
\[
P(k) \propto e^{-\lambda k},
\]
where $\lambda$ is the decay rate. 

As $\Delta T$ increases, the decay rate $\lambda$ decreases, reflecting a broader degree distribution. This means that longer observation windows ($\Delta T = 7$) capture more medium- and high-degree nodes than shorter windows ($\Delta T = 1$). This transition from localized interactions to broader connectivity aligns with the increasing temporal scope, which aggregates more co-visitation patterns.

Figure \ref{fig:structural_properties}B illustrates the hyperedge size distribution across $\Delta T \in \{1, 3, 7\}$. The hyperedge size, defined as the number of nodes in a hyperedge, is plotted on the x-axis, with the y-axis showing the frequency in log-scale. The maximum hyperedge size increases from 4 for $\Delta T = 1$ to 5 for $\Delta T = 3$, and 7 for $\Delta T = 7$, emphasizing the growth of higher-order interactions as $\Delta T$ increases. Notably, larger hyperedges are more prevalent in weekly windows ($\Delta T = 7$), indicating that longer temporal windows facilitate higher-order interactions.

Figure \ref{fig:structural_properties}C examines the maximum chebyshev distance among nodes in hyperedges of size at least 3 (i.e., only in higher-order interactions). The plot shows that for $min\_sup = 0.005$, the maximum distance increases sharply from $\Delta T = 3$ to $\Delta T = 7$, indicating more spatially distributed interactions. For higher $min\_sup$ values, the increase is greater from $\Delta T = 1$ to $\Delta T = 3$, reflecting a reduced presence of high-support interactions occurring over large distances.

Figure \ref{fig:structural_properties}D-F displays spatial heatmaps of node degrees on a 20x20 grid for $min\_sup = 0.005$ and $\Delta T \in \{1, 3, 7\}$. Each node's position corresponds to its $(x, y)$ coordinates, with the intensity of the red shading indicating its degree. As $\Delta T$ increases from 1 to 7, the spatial distribution of hyperedges becomes more diffuse, covering larger regions of the grid. The maximum node degree increases significantly, from 44 for $\Delta T = 1$, to 502 for $\Delta T = 3$, and to 7402 for $\Delta T = 7$, reflecting an increasing aggregation of nodes into hyperedges over longer temporal windows. Central high-degree nodes contribute to more hyperedges as $\Delta T$ increases, emphasizing their role in broader co-visitation patterns. Given the city-scale analysis, few locations are naturally highly connected, as reflected in the heatmaps. 

\subsection{Structural and Spatial Comparisons of Co-Visitation Patterns Between Regular and Emergency Days}
\begin{figure}
    \centering
    \includegraphics[scale=0.6]{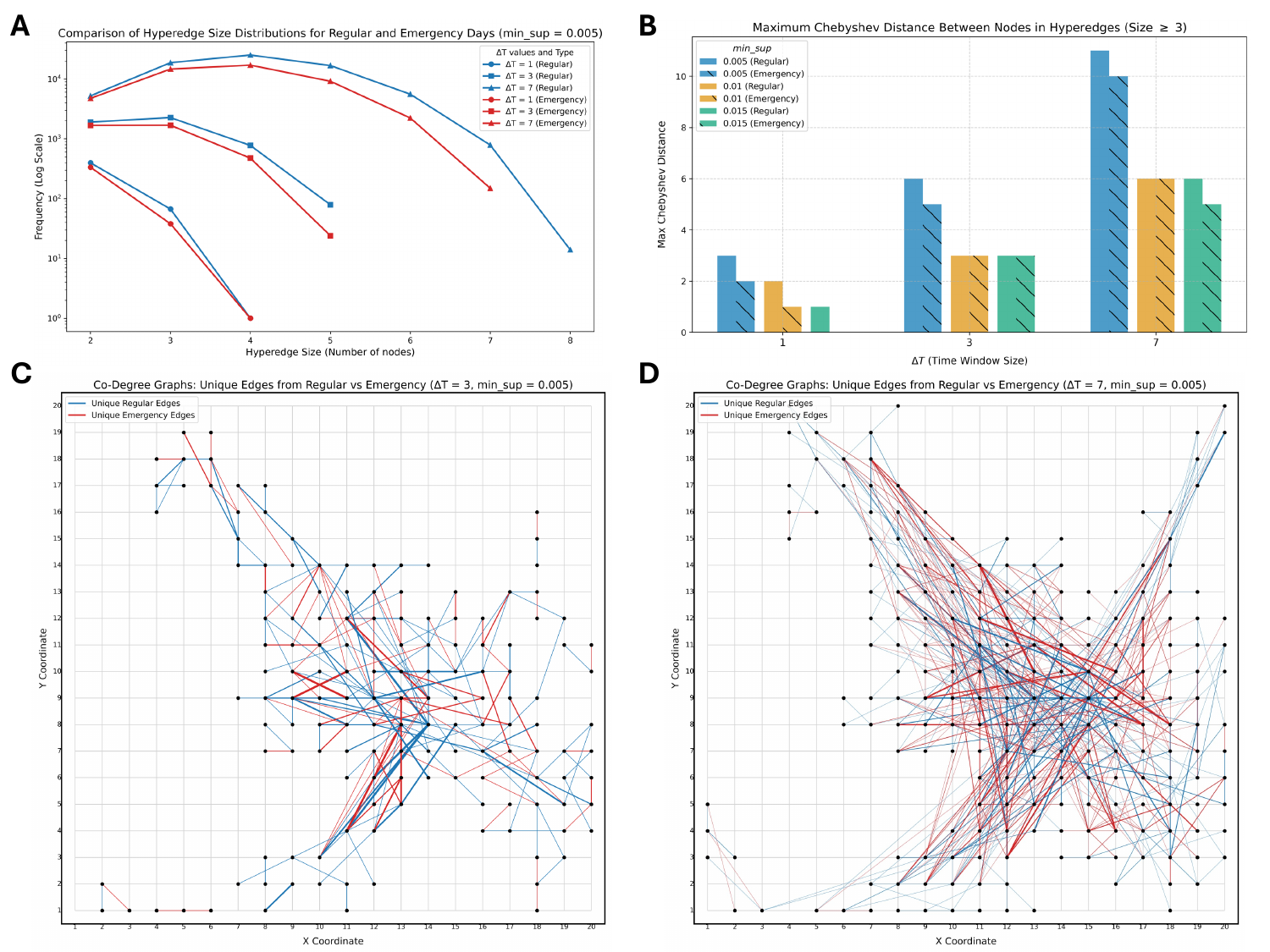}
    \caption{
    \textit{Comparison of hypergraphs capturing mobility patterns on regular and emergency days.}
    \textbf{A}) Hyperedge size distribution for $\Delta T$ values at $min\_sup = 0.005$. The x-axis shows hyperedge size, and the y-axis (log scale) represents frequency. Blue curves denote regular days, red curves emergency days, with markers for $\Delta T$. Larger hyperedges are more frequent on regular days, with differences increasing at higher $\Delta T$.  
    \textbf{B}) Maximum Chebyshev distance for higher-order interactions (hyperedges $\geq 3$) across $\Delta T$ and $min\_sup$. Solid bars show regular days, striped bars emergency days, with colors for $min\_sup$ values. Differences peak at $min\_sup = 0.005$, suggesting reduced long-distance movement during emergencies. 
    \textbf{C--D}) Unique edges in co-degree graphs for regular and emergency hypergraphs at $\Delta T = 3$, $min\_sup = 0.005$ and  $\Delta T = 7$, $min\_sup = 0.005$. Blue edges are unique to regular days, while red edges are unique to emergency days. Edge thickness reflects co-occurrence frequency. These visualizations highlight shifts in co-visitation patterns during emergencies.
    }
    \label{fig:regvsemg}
\end{figure}

Figure \ref{fig:regvsemg}A presents the hyperedge size distribution for regular and emergency days at $min\_sup = 0.005$. The x-axis represents hyperedge size, while the y-axis, plotted on a logarithmic scale, shows the frequency of hyperedges of that size. Regular-day hypergraphs (blue curves) exhibit a higher frequency of larger hyperedges compared to emergency-day hypergraphs (red curves). This difference is particularly pronounced for larger $\Delta T$ values, such as 7, where the longer observation window allows for greater aggregation of co-visitation patterns.

The maximum Chebyshev distance computed over all hyperedges of size $\geq 3$ analyzed in Figure \ref{fig:regvsemg}B  captures the largest spatial extent of higher-order interactions in a hypergraph. For regular days (solid bars) and emergency days (striped bars), the chebyshev distance increases with $\Delta T$ and decreases as $min\_sup$ increases. The largest differences between regular and emergency days are observed at $min\_sup = 0.005$, where the chebyshev distance during emergencies is smaller by 1 consistently over all time window sizes. This reduction highlights the restricted spatial spread of higher-order interactions during emergencies, consistent with localized mobility patterns observed in Figure \ref{fig:regvsemg}A.

Figures \ref{fig:regvsemg}C visualizes the unique edges in co-degree graphs for regular and emergency hypergraphs. A co-degree graph represents the relationships between node pairs based on their co-occurrence in hyperedges. Formally, the co-degree of a pair of nodes $(u, v)$ is the number of hyperedges in which both nodes appear together~\cite{Alon2005-zu}. The edge weight in a co-degree graph corresponds to this co-degree value, effectively summarizing pairwise interaction strength. We particularly restrict our co-degree graphs to focus on the hyperedges of size $\geq 3$, thereby restricting the analysis to higher-order interactions. In Figures \ref{fig:regvsemg}C-D, blue edges represent co-degree relationships unique to regular days, while red edges indicate those unique to emergency days. The thickness of each edge corresponds to its co-degree value, with thicker edges indicating stronger pairwise interactions. Edges common to both regular and emergency days are excluded from the plots. As $\Delta T$ increases to 7 (Figure \ref{fig:regvsemg}D), the number of unique edges increases, extending over a broader spatial range. This analysis provides insights into locations that are co-visited exclusively during emergencies or remain popular during regular days but see reduced or no co-visitation in emergency scenarios.

The hyperedge size distributions and co-degree graph visualizations underscore key differences in mobility patterns between regular and emergency days. Regular-day hypergraphs exhibit broader spatial coverage, more higher-order interactions, particularly for larger $\Delta T$.

\subsection{Analysis of POI Distribution and its Relationship with Node Degree in Hypergraphs}
\begin{figure}
    \centering
    \includegraphics[scale=0.46]{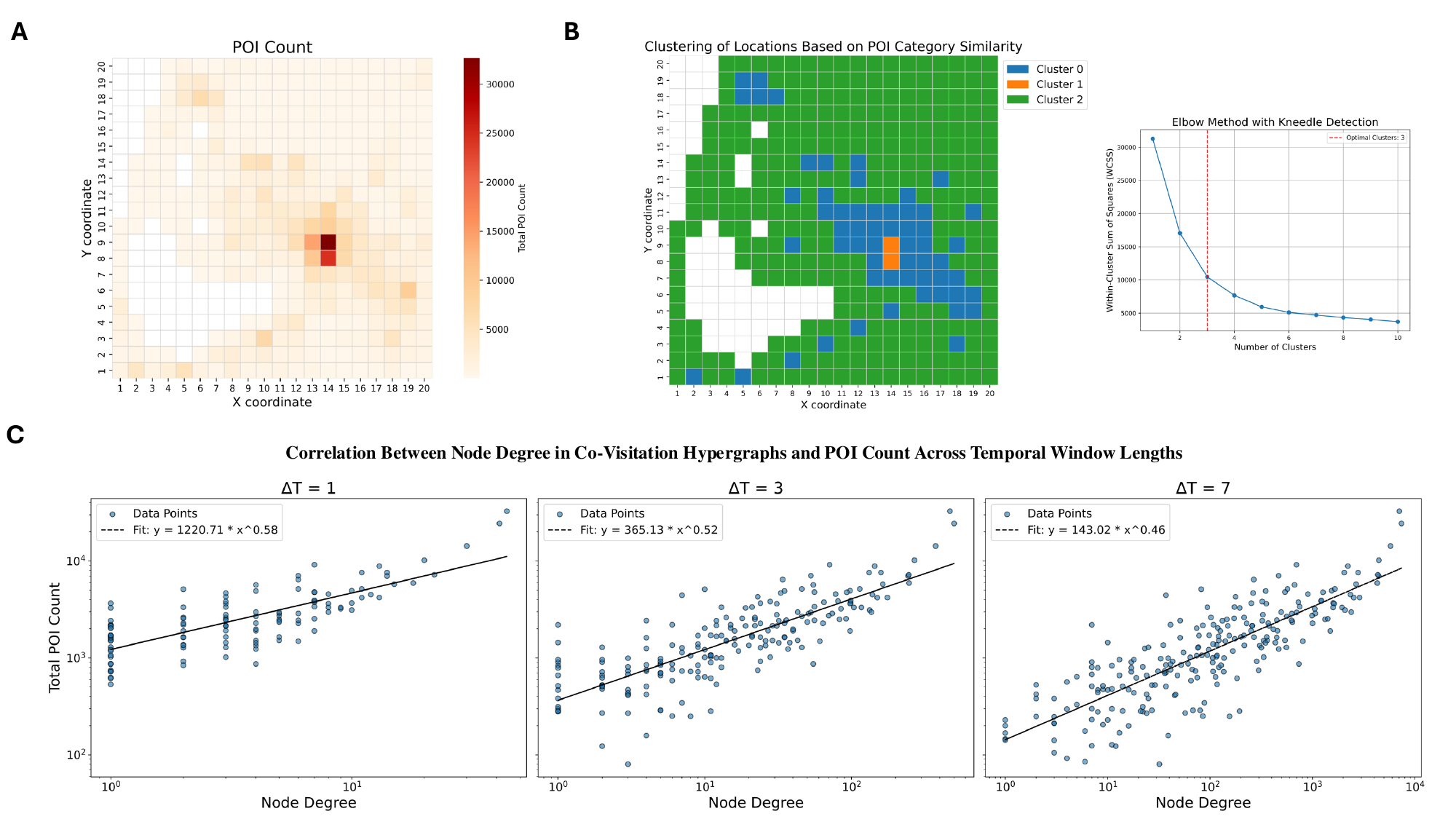}
    \caption{\textit{POI Distribution Among Locations and Its Relationship with Degree Centrality in Hypergraphs}. 
\textbf{A}) Heatmap showing the total number of Points of Interest (POI) at each location, represented by X and Y coordinates. Darker shades of red indicate higher POI counts.
\textbf{B}) Clustering of locations obtained using K-means clustering. Each location is represented by a vector corresponding to the number of POIs of each category. The similarity between these POI vectors is used as the metric, and colors represent cluster membership. 
\textbf{C}) Scatter plots depicting the correlation between the total POI count at each location and the corresponding degree centrality in hypergraphs generated across varying temporal window lengths ($\Delta T$). The x-axis represents the degree of a location in hypergraph for given $\Delta T$ and $min\_sup=0.015$, while the y-axis represents the total POI count in that location. The dashed black line represents the best power-law fit to the data.}
    \label{fig:poi_vs_degree}
\end{figure}

The spatial distribution of Points of Interest (POI) plays a crucial role in shaping mobility patterns, as locations with higher POI density have higher gravity and often attract greater foot traffic~\cite{Simini2021-cj}. Figure \ref{fig:poi_vs_degree} provides insights into the distribution of POIs across the spatial grid, their clustering patterns, and their relationship with degree centrality in co-visitation hypergraphs.

\paragraph{Spatial Distribution of POIs.} 
The heatmap in Figure \ref{fig:poi_vs_degree}A reveals significant heterogeneity in the distribution of POIs across the 20x20 grid. A few locations exhibit a markedly higher density of POIs, as indicated by darker shades of red, suggesting that these areas may serve as hubs of activity within the city. These nodes contribute disproportionately to the overall mobility network by attracting individuals for various purposes, ranging from work to leisure. 

\paragraph{Clustering of Locations Based on POI Categories.}
To explore patterns in POI composition, Figure \ref{fig:poi_vs_degree}B presents clusters of locations derived using K-means clustering. Each location is represented as a vector containing the counts of POIs in various categories (e.g., retail, healthcare, and education). By using the similarity of these POI vectors as the clustering metric, three distinct clusters are identified, with the optimal number of clusters determined via the Elbow method and kneedle detection\cite{Satopaa2011-ll} to minimize Within-Cluster Sum of Squares (WCSS). We find that the optimal number of clusters is just 3. A limitation to finding more nuanced clustering is the fact that the categories are anonymized in this dataset for privacy purposes.

\paragraph{Relationship Between POI Density and Node Degree.}
Figure \ref{fig:poi_vs_degree}C examines the correlation between the total POI count at each location and its degree in hypergraphs constructed for varying temporal window lengths ($\Delta T$) with $min\_sup = 0.015$. Degree centrality represents the total number of hyperedges incident on a node, capturing the extent of co-visitation interactions associated with the location. The scatter plot indicates a positive correlation between POI count and node degree, suggesting that locations with higher POI density tend to play a more central role in the co-visitation network.

The best-fit parameters for the power-law model of the form $y = a \cdot x^b$, where $y$ represents the node degree and $x$ represents the total POI count, are estimated. The dashed black line in Figure \ref{fig:poi_vs_degree}C visualizes the fit. The observed power-law relationship highlights the disproportionate influence of high-POI locations, which act as hubs in the mobility network. Overall, this analysis underscores the strong interplay between urban structure, as reflected by POI distributions, and the structural properties of mobility hypergraphs. 

%% file: discussion.tex
\section{Discussion}
In this paper, we applied our approach to a city-scale mobility dataset and revealed the structure and spatial properties of co-visitation hypergraphs in this specific context. However, our methods could be applied to inter-city or regional mobility data to reveal broader co-visitation patterns. Further, the comparison of flow networks with co-visitation hypergraphs has potential to highlight scenarios where hypergraphs offer deeper insights. Applying this approach on longer temporal datasets could enhance understanding of seasonal and long-term mobility trends.

\subsection{Potential Applications}
The co-visitation hypergraph framework is valuable for analyzing group-level interactions, particularly in applications like contact tracing in infectious disease epidemiology. 

During epidemics, hypergraphs can identify higher-order movement patterns driving disease spread, supporting targeted interventions like optimized testing or localized travel restrictions. Contact tracing would normally collect the information about places an infected individual visited, but the hypergraph approach allows to estimate the plausible reach of individuals at these locations. This reach can be tailored to epidemiologically meaningful frameworks (i.e., using a $\Delta T$ concordant with infection progression). For contact tracing, analyzing mobility within a defined time window (e.g., three days) helps track all locations visited by exposed individuals. If infections emerge at a site, co-visitation hypergraphs help reveal all connected locations within the time window, facilitating efficient surveillance and containment.

In climate emergencies, hypergraphs capture shifts in mobility behaviors during crises. Certain movement patterns (hyperedges) may emerge or disappear, which traditional mobility networks often miss. By offering a more comprehensive view of movement adaptations, hypergraphs enhance evacuation planning, resource allocation, and post-disaster recovery by identifying critical travel routes and behavioral shifts.

\subsection{Limitations of the Current Study}
Our approach to constructing mobility hypergraphs relies on individual-level mobility data, which limits its applicability to datasets with such granularity. The datasets used in our study span only 75 days, with emergency scenarios covering just 15 days in DS2. Longer time-series data would allow for a more comprehensive study of temporal evolution in co-visitation patterns, particularly during emergencies. Furthermore, the datasets anonymize location names and Points of Interest (POIs), which restricts the ability to conduct application-oriented analyses such as understanding the role of specific types of locations (e.g., schools, hospitals, markets) in mobility dynamics. Finally, our study is constrained to city-scale data; incorporating datasets with larger spatial coverage, such as inter-city or regional mobility, could provide insights into how structural properties of hypergraphs vary across broader geographic scales.

%% file: conlusions.tex
\section{Conclusions}
We introduce a novel representation of mobility data that captures higher-order interactions among locations. We propose a method for constructing co-visitation hypergraphs using individual-level mobility trajectory data. Our results validate the utility of hypergraphs in capturing complex, higher-order mobility patterns, making them valuable for urban mobility analysis, public health, and disaster management. In conclusion, our hypergraph-based mobility analysis framework advances research in urban planning, public health, and disaster resilience, opening new avenues for future applications.

%% file: appendix.tex
\section*{\centering \LARGE Appendix}

\section{Limitations of Existing Location-to-Location Interactions Based Representations of Human Mobility}
\begin{figure}[H]
    \centering
    \includegraphics[height=3.5cm, width = 12cm]{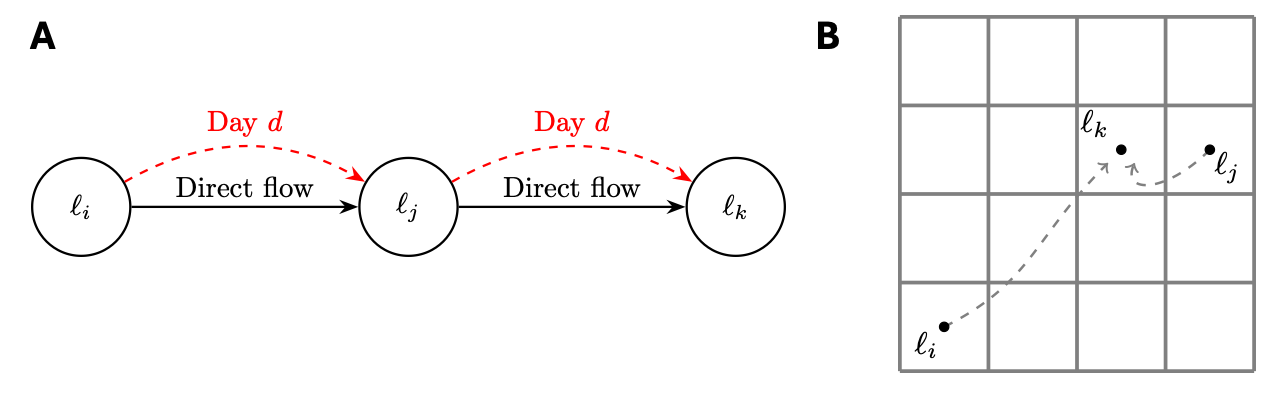}
    \caption{\textit{Location-to-Locations Interactions in mobility flow networks and colocation maps.} \textbf{A}) The pairwise interactions in a flow network, $(\ell_i, \ell_j)$ and $(\ell_j, \ell_k)$, do not confirm the existence of a path $\ell_i \to \ell_j \to \ell_k$ (shown in red dashed lines) traveled by individuals within the same time interval (e.g., on the same day). Additionally, they do not indicate whether the locations $\{\ell_i, \ell_j, \ell_k\}$ will be visited together on a single day. \textbf{B}) The figure shows a colocation event where individuals from locations $\ell_i$ and $\ell_j$ are in the same place at the same time, captured as a pairwise interaction in colocation maps. }
    \label{fig:flownet_implication}
\end{figure}

A mobility flow network $G = (V, E)$ is a weighted directed graph where nodes in $V$ represent locations, edges in $E$ indicate movement of individuals between location pairs, and edge weights quantify the intensity of that movement. For example, consider the flow between locations $(\ell_i, \ell_k) \not \in E$ (locations without a direct edge in the flow network $G$), we need to consider the paths connecting them. For instance, if $P = \{\ell_i, \ell_j, \ell_k\}$ is a path in $G$ where $(\ell_i, \ell_j)$ and $(\ell_j, \ell_k)$ are edges in $E$, this indicates movement between $\ell_i$ and $\ell_j$, as well as between $\ell_j$ and $\ell_k$. However, this does not guarantee that some individuals actually traveled in the exact sequence $\ell_i \to \ell_j \to \ell_k$ or even that they visited all three locations together in a time interval. In other words, the existence of such a path does not imply that any person followed that specific route. This is illustrated in Figure \ref{fig:flownet_implication}A. The mobility flow network representation also overlooks complex interactions, like flow that includes more than two locations, which could offer more detailed insights into human mobility patterns.

Colocation maps estimate the likelihood that individuals from different home locations are present in the same place at the same time, as illustrated by a colocation event in Figure \ref{fig:flownet_implication}B. These maps reveal how populations from various locations come into contact, emphasizing the spatial and temporal overlap of individuals within a given time window. However, this approach is limited to pairwise relationships between locations and does not account for the dynamic interactions or movement flows between them.

In this work, we focus on higher-order flows among locations, capturing frequent mobility patterns within a region. Humans visit sequences of locations within a given time window (e.g. a day or during a week, weekend), forming trajectories on a map. For example, we examine the proportion of individuals in a population visiting three locations—$\ell_i$, $\ell_j$, and $\ell_k$—in any order during a specified period.

Building on this concept, we generalize mobility networks by representing frequent mobility patterns as higher-order interactions between locations, encoded as hyperedges in a mobility hypergraph. This hypergraph-based approach offers a versatile framework for analyzing and applying mobility flows in various contexts.

Unlike pairwise centrality in traditional networks, hypergraph centrality can identify not just key nodes (locations) but also pivotal hyperedges (groups of locations) that are central to movement patterns. This is particularly useful for pinpointing influential areas in disease spread or critical hubs during emergencies, providing actionable insights for intervention. By integrating spatial, temporal, and structural dimensions, mobility hypergraphs offer a powerful tool for comprehensive analysis and informed decision-making.
\section{Undirected vs. Directed Mobility Hypergraphs}

In the mobility hypergraph \(\mathbb{H}\), each hyperedge represents an interaction between two or more locations. The edges are typically undirected, as in many applications, the sequence of visits to these locations may not be relevant. For instance, in contact tracing for epidemic control, it is often sufficient to identify the set of locations an individual has visited, regardless of the order in which those visits occurred. However, in other contexts, the order of location visits may have significant importance. In such cases, the same approach can be extended to construct a directed hypergraph\cite{directedhypergraph2017}, where the hyperedges (hyperarcs) represent ordered sequences of locations rather than unordered sets.

\section{Impact of $min\_sup$ and $\Delta T$ on the Evolution of Co-Visitation Hypergraphs}

\begin{figure*}
    \centering
    \includegraphics[height=12cm, width = 16cm]{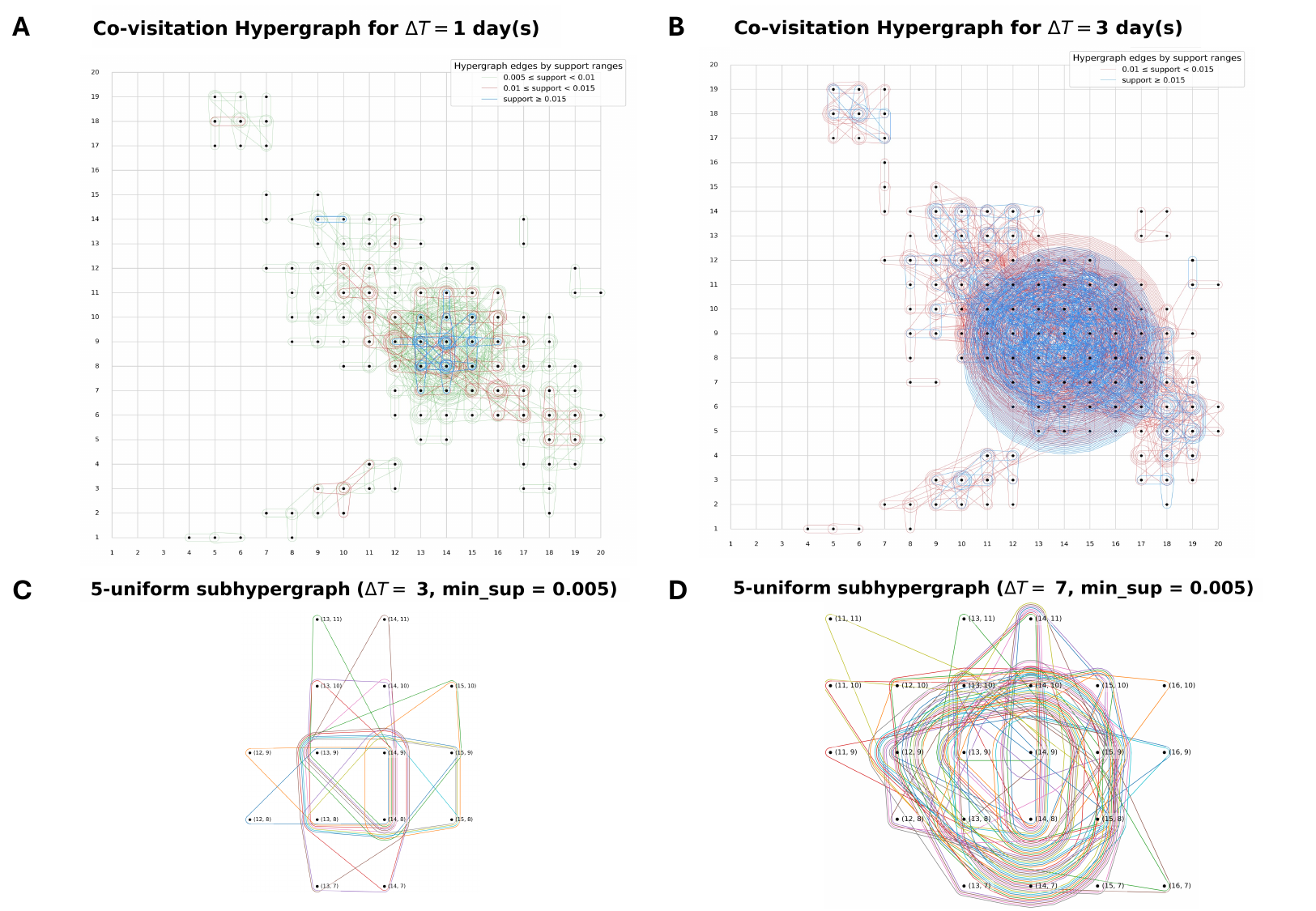}
    \caption{\textit{Visualization of Co-Visitation Hypergraphs and k-Uniform Subhypergraphs.}  
\textbf{A}) Co-visitation hypergraph for $\Delta T = 1$ and $min\_sup = 0.005$, with nodes shown as black dots on a 20x20 spatial grid and hyperedges represented as elastic bands. Hyperedge colors indicate support ranges: green for $[0.005, 0.01)$, red for $[0.01, 0.015)$, and blue for $[0.015, \infty)$. The visualisations are generated using the HyperNetX package developed by Pacific Northwest National Laboratory~\cite{Praggastis2024-uo}.
\textbf{B}) Co-visitation hypergraph for $\Delta T = 3$ and $min\_sup = 0.01$, depicted similarly to A, with hyperedge colors limited to red and blue for corresponding support ranges.  
\textbf{C-D}) 5-uniform subhypergraphs of co-visitation hypergraphs for $\Delta T = 3$ and $\Delta T = 7$ with $min\_sup = 0.005$. Nodes, shown as black dots, are positioned on a cropped spatial grid containing only subhypergraph nodes. All hyperedges are of size 5. The maximum Chebyshev distance between connected nodes is 3 for $\Delta T = 3$ and 4 for $\Delta T = 7$. These visualizations highlight variations in topology, spatial extent, edge density, and spatial proximity of nodes in co-visitation hypergraphs with changes in $\Delta T$ and $min\_sup$.  
}
    \label{fig:impact_deltaT_minsup}
\end{figure*}

We explore how the time window size ($\Delta T$) and the minimum support threshold ($min\_sup$) influence the evolution of co-visitation hypergraphs, focusing on their structural and spatial characteristics. Figures \ref{fig:impact_deltaT_minsup}A-B illustrate co-visitation hypergraphs for daily ($\Delta T = 1$) and three-day ($\Delta T = 3$) observation windows, respectively. Nodes are positioned on a 20x20 spatial grid based on their coordinates, and hyperedges are visualized as elastic bands connecting incident nodes. The hyperedge colors represent their support levels: green ($[0.005, 0.01)$), red ($[0.01, 0.015)$), and blue ($[0.015, \infty)$). For a fixed $min\_sup$ threshold, increasing $\Delta T$ results in a significant rise in the number and spatial coverage of hyperedges. For example, for $min\_sup = 0.01$ (corresponding to 1\%), the hyperedges (shown in blue) expand from being concentrated around high-degree nodes for $\Delta T = 1$ to a larger spatial distribution for $\Delta T = 3$. This reflects how a longer temporal window captures a more diverse set of interactions across the spatial grid.

Figures \ref{fig:impact_deltaT_minsup}C-D provide visualizations of 5-uniform subhypergraphs derived from the co-visitation hypergraphs for $\Delta T = 3$ and $\Delta T = 7$, respectively, with $min\_sup = 0.005$. These subhypergraphs contain only hyperedges of size 5. The maximum Chebyshev distance between nodes in each hyperedge is calculated to examine the spatial proximity of higher-order interactions. This is applicable in our context only because our nodes form grid cells in a rectangular grid. The Chebyshev distance\cite{De_Oliveira2020-ay} between two nodes at coordinates $(x_1, y_1)$ and $(x_2, y_2)$ is defined as:  

\[
d_\infty((x_1, y_1), (x_2, y_2)) = \max(|x_1 - x_2|, |y_1 - y_2|).
\]  

\noindent
For each hyperedge $e$, the maximum Chebyshev distance is given by:  
\[
d_\infty(e) = \max_{u, v \in e} d_\infty(u, v),
\]  
where $u$ and $v$ are any two nodes in the hyperedge $e$. Therefore, we define a hypergraph $\mathbb{H} = (V, \mathcal{E)}$, the maximum chebyshev distance is $D_\infty(\mathcal{H}) = \max_{e \in \mathcal{E}} d_\infty(e)$. For $\Delta T = 3$, the maximum Chebyshev distance in the 5-uniform subhypergraph is 3, while for $\Delta T = 7$, it increases to 4. This demonstrates how larger temporal windows not only lead to more higher-order interactions but also allow interactions to occur over greater spatial distances.